\documentclass[11pt,a4paper]{llncs}
\usepackage{graphicx}
\usepackage{amsmath,amssymb,mathrsfs}

\def\Tr{\operatorname{Tr}} 
\def\>{\rangle}\def\<{\langle} \def\sH{\mathscr{H}}

\def\states{\mathcal{S}}
\def\N#1{|\!|{#1}|\!|}

\def\mR{\mathcal{R}} 
\def\id{\operatorname{id}} 
\def\mN{\mathcal{N}} \def\fg{\operatorname{g}}

\def\qcb{Q_{cb}}
\DeclareRobustCommand\openone{\leavevmode\hbox{\small1\normalsize\kern-.33em1}}

\begin{document}
\title{Irreversibility of entanglement loss}
\author{Francesco Buscemi}
\institute{ERATO-SORST Quantum Computation and Information Project,\\
  Japan Science and Technology Agency,\\
e-mail: \texttt{buscemi@qci.jst.go.jp}}

\maketitle       
\begin{abstract}
  The action of a channel on a quantum system, when non trivial,
  always causes deterioration of initial quantum resources, understood
  as the entanglement initially shared by the input system with some
  reference purifying it. One effective way to measure such a
  deterioration is by measuring the loss of coherent information,
  namely the difference between the initial coherent information and
  the final one: such a difference is ``small'', if and only if the
  action of the channel can be ``almost perfectly'' corrected with
  probability one.

  In this work, we generalise this result to different entanglement
  loss functions, notably including the entanglement of formation
  loss, and prove that many inequivalent entanglement measures lead to
  equivalent conditions for approximate quantum error correction. In
  doing this, we show how different measures of bipartite entanglement
  give rise to corresponding distance-like functions between quantum
  channels, and we investigate how these induced distances are related
  to the cb-norm.
\end{abstract}

\section{Introduction}

What is irreversibility of a process? This question, in this form,
does not make much sense. We first have to specify ``irreversibility
with respect to what''. It means we first need to decide a set of
rules---i.~e. a set of allowed transformations together with some free
resource---to which one has to conform when trying to revert the
process. We can then say that irreversibility basically measures the
deterioration of some resource that does not come for free, within the
rules we specified. When studying quantum error correction, one
usually considers an extremely strict scenario, where legitimate
corrections only amount to a fixed quantum channel applied after the
action of the noise\footnote{This is different, for example, from the
  correction of quantum measurements~\cite{ours}: in such a case, we
  can access classical information produced by the measurement
  apparatus. Therefore, in general, it is easier (in the sense that
  the set of allowed transformations is larger) to correct quantum
  measurements than quantum channels. Another case is that of
  environment assisted quantum error correction, where we are allowed
  not only to access classical information from the environment, but
  we can also choose the measurement to perform onto
  it~\cite{env-ass}.}. This scenario corresponds to the task of trying
to restore the entanglement initially shared by the input system
(undergoing the noise) with an inaccessible reference, only by using
local actions on the output system, being any kind of communication
between the two systems impossible.

Being quantum error correction a basic task in quantum information
theory, the literature on the subject grew rapidly in the last 15
years~\cite{qec}. It is however possible to devise two main sectors of
research: the first one is devoted to the design of good quantum error
correcting codes, and directly stems from an algebraic approach to
\emph{perfect} quantum error correction; the second one tries to
understand conditions under which \emph{approximate} quantum error
correction is possible. Usually, while the former is more practically
oriented, the latter is able to give information theoretical bounds on
the performance of the optimum correction strategy, even when perfect
correction is not possible, while leaving unspecified the optimum
correction scheme itself.

Our contribution follows the second approach: we will derive some
bounds relating the loss of entanglement due to the local action of a
noisy channel on a bipartite state with the possibility of undoing
such a noise. The original point in our analysis is that we will
consider many inequivalent ways to measure entanglement in bipartite
mixed states, hence obtaining many inequivalent measures of
irreversibility. After reviewing the main results of
Ref.~\cite{buscemi}, we will show how we can relate such entropic
quantities with different norm-induced measures of irreversibility,
like those exploiting the cb-norm distance~\cite{dennis} or the
channel fidelity~\cite{mauro-bel-rag}, therefore providing measures of
the overall---i.~e. state independent---irreversibility of a quantum
channel.

\section{Evaluating the coherence of an evolution}

In the following, quantum systems will be often identified with the
(finite dimensional) Hilbert spaces supporting them, that is, the
roman letter $A$ [resp. $B$], rigorously denoting the system only,
will also serve as a shorthand notation instead of the more explicit
$\sH^A$ [resp. $\sH^B$]. The (complex) dimension of $\sH^A$
[resp. $\sH^B$] will be denoted as $d_A$ [resp. $d_B$]. The set of
possible states of the system $A$ [resp. $B$], that is, the set of
positive semi-definite operators with unit trace acting on $\sH^A$
[resp. $\sH^B$], will be equivalently denoted with $\states(\sH^A)$
[resp. $\states(\sH^B)$] or $\states(A)$ [resp. $\states(B)$].

A general quantum noise $\mN:\states(A)\to\states(B)$ is described as
a completely positive trace-preserving map---i.~e. a
\emph{channel}. If the input system $A$ is initially described by the
state $\rho^A$, we will write $\sigma^B$ to denote $\mN(\rho^A)$. The
aim of this section is to understand how one can measure the coherence
of the evolution
\begin{equation}\label{eq:evolution}
  \rho^A\mapsto\sigma^B:=\mN(\rho^A)
\end{equation}
induced by $\mN$ on $\rho^A$. (We will see in the following how to get
rid of the explicit dependence on the input state and obtain a
quantity measuring the overall invertibility of a given channel, as a
function the channel only.)

\begin{figure}[h,c]
\begin{center}
  \includegraphics[width=8cm]{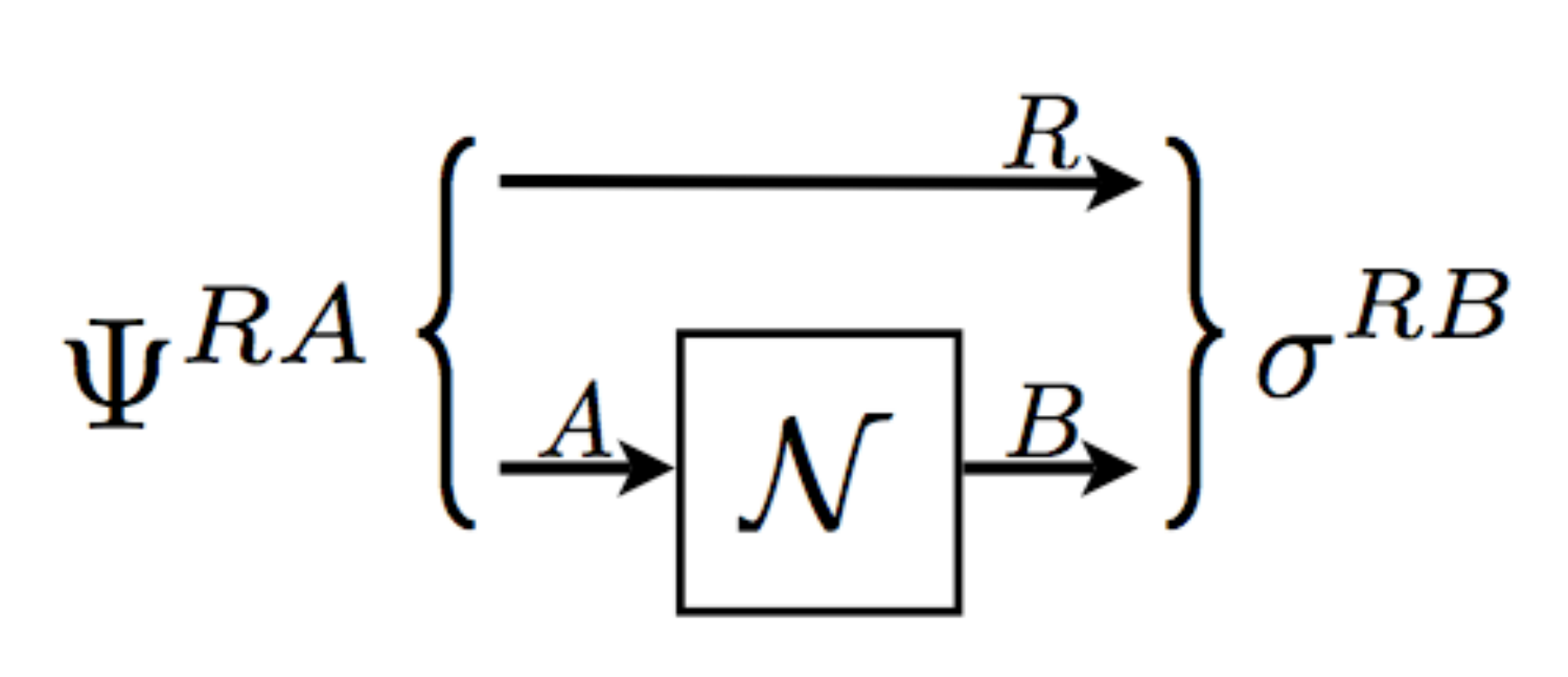}\end{center}
\caption{The input state $\rho^A$ is purified with respect to a
  reference system $R$ into the state $|\Psi^{RA}\>$. The noise
  $\mN:A\to B$ acts on the system $A$ only, in such a way that
  $|\Psi^{RA}\>$ is mapped into
  $\sigma^{RB}:=(\id^R\otimes\mN^A)(\Psi^{RA})$.}
  \label{fig:3}
\end{figure}

Before continuing the discussion, we should clarify what we mean with
the term ``coherence''. Imagine that the input system $A$ is actually
the subsystem of a larger bipartite system $RA$, where the letter $R$
stands for \emph{reference}, initially described by a pure state
$|\Psi^{RA}\>$, such that
\begin{equation}
\Tr_R[\Psi^{RA}]=\rho^A.
\end{equation}
The situation is depicted in Fig.~\ref{fig:3}.
Notice that the input state
$\rho^A$ is mixed if and only if the pure state $|\Psi^{RA}\>$ is
entangled. Then, the coherence of the evolution~(\ref{eq:evolution})
can be understood as the amount of residual entanglement survived in
the bipartite output (generally mixed) state
$\sigma^{RB}:=(\id^R\otimes\mN^A)(\Psi^{RA})$ after the noise locally
acted on $A$ only. However, any naive attempt to formalise such an
intuitive idea is soon frustrated by the fact that there exist many
different and generally inequivalent ways to measure the entanglement
of a mixed bipartite
system~\cite{christhesis,hayashi,generic_ent}. This well-known
phenomenon turns out in the existence of many different and generally
inequivalent, but all in principle valid, ways to measure the
coherence of an evolution.

\begin{figure}[h,c]
\begin{center}
  \includegraphics[width=10cm]{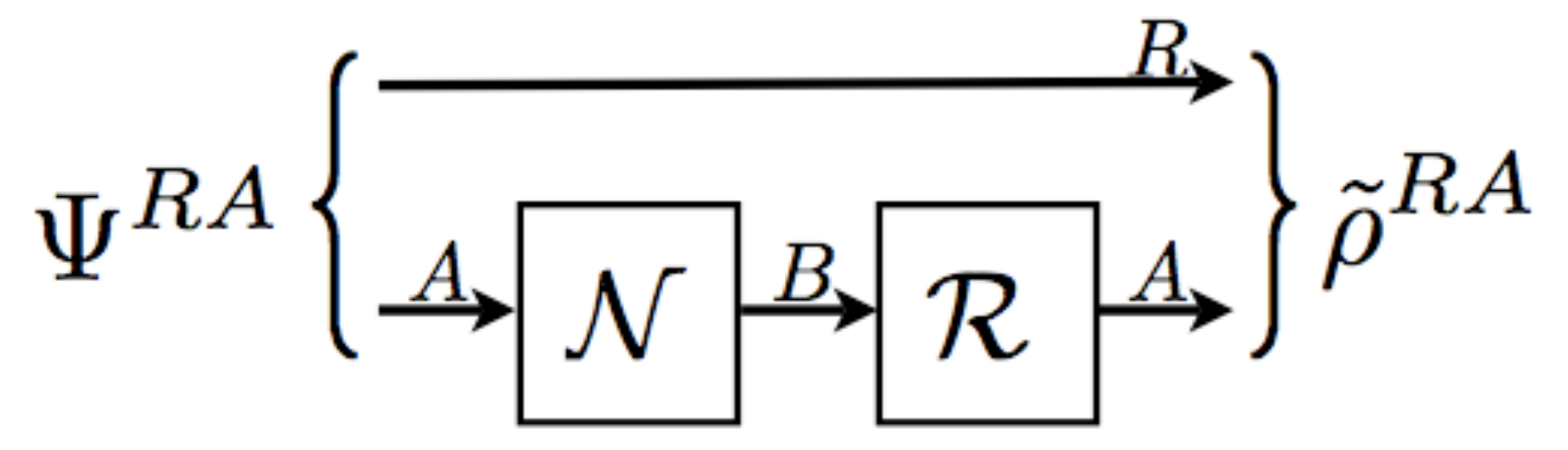}\end{center}
\caption{With respect to Fig.~\ref{fig:3}, here, after the noise
  $\mN$, we apply a subsequent correction via a local restoring
  channel $\mR:B\to A$. The corrected bipartite output state
  $(\id^R\otimes\mR^B\circ\mN^A)(\Psi^{RA})$ is denoted by
  $\tilde\rho^{RA}$.}
  \label{fig:1}
\end{figure}

One possibility to overcome such a problem was considered already in
Ref.~\cite{schum}. There, Schumacher introduced the quantity called
\emph{entanglement fidelity} of a channel $\mN:A\to A$ with respect to
an input state $\rho^A$, defined as
\begin{equation}\label{eq:ent_fid}
F_e(\rho^A,\mN):=\<\Psi^{RA}|(\id^R\otimes\mN^A)(\Psi^{RA})|\Psi^{RA}\>.
\end{equation}
Such a quantity (which does not depend on the particular purification
$|\Psi^{RA}\>$ considered) accurately describes how close the channel
$\mN$ is to the noiseless channel $\id$ on the support of
$\rho^A$~\cite{schum}. However, it was noticed that, as defined in
Eq.~(\ref{eq:ent_fid}), $F_e(\rho^A,\mN)$ is \emph{not} related to the
coherence of the evolution $\rho^A\mapsto\mN(\rho^A)$, in that it is
easy to see that a unitary channel---i.~e. completely coherent---can
result in a null entanglement fidelity. We then have to consider a
more general situation, like the one depicted in Fig.~\ref{fig:1}.
After the local noise produced the bipartite state
$\sigma^{RB}$, we apply a local restoring channel $\mR:B\to A$ to
obtain
\begin{equation}
\tilde\rho^{RA}:=(\id^R\otimes\mR^B\circ\mN^A)(\Psi^{RA}).
\end{equation}
Notice that in general the restoring channel can explicitly depend on
the input state $\rho^A$ and on the noise $\mN$. However, for sake of
clarity of notation, we will leave such dependence understood, and
make it explicit again, by writing $\mR_{\rho,\mN}^B$, only when
needed. We now compute the \emph{corrected} entanglement fidelity
$F_e(\rho^A,\mR\circ\mN)$ and take the supremum over all possible
corrections
\begin{equation}\label{eq:corr_ent_fid}
  \overline{F}_e(\rho^A,\mN):=\sup_{\mR_{\rho,\mN}}F_e(\rho^A,\mR_{\rho,\mN}\circ\mN).
\end{equation}
This is now a good measure of the coherence of the noisy evolution
$\rho^A\mapsto\mN(\rho^A)$: by construction it is directly related
to the degree of invertibility of the noise $\mN$ on the support of
$\rho^A$.

\section{Coherent information loss}

The maximisation over all possible correcting channels in
Eq.~(\ref{eq:corr_ent_fid}) can be extremely hard to
compute. Moreover, we are still interested in understanding how the
coherence of a transformation is related to the theory of bipartite
entanglement. The idea is that of finding some quantity (typically an
entropic-like function) which is able to capture at one time both the
amount of coherence preserved by the channel as well as the
invertibility of the channel itself, possibly bypassing the explicit
evaluation of $\overline{F}_e(\rho^A,\mN)$, for which accurate upper
and lower bounds would suffice.

A key-concept in the theory of approximate quantum error correction is
that of \emph{coherent information}~\cite{schum,lloyd}, which, for a
bipartite state $\tau^{AB}$, is defined as
\begin{equation}
  I_c^{A\to B}:=S(\tau^B)-S(\tau^{AB}),
\end{equation}
where $S(\tau):=-\Tr[\tau\log_2\tau]$ is the von Neumann entropy of
the state $\tau$. Notice that, in the definition of coherent
information, system $A$ and system $B$ play apparently different
roles: such asymmetry acknowledges that the flow of quantum
information is considered as being from $A$ to $B$. Accordingly,
\emph{channel coherent information} is defined as
\begin{equation}
  I_c(\rho^A,\mN):=I_c^{R\to B}(\sigma^{RB})=S(\sigma^{B})-S(\sigma^{RB}),
\end{equation}
where $R,A,B$ stand for reference, input, and output system,
respectively. In our picture, the input state $|\Psi^{RA}\>$ is pure,
so that $I_c^{R\to A}(\Psi^{RA})=S(\rho^A)=S(\rho^R)$. We then compute
the coherent information loss due to the action of the noise $\mN$ on
subsystem $A$ as
\begin{equation}\label{eq:coherent-info-loss}
\begin{split}
  \delta_c(\rho^A,\mN)&:=I_c^{R\to A}(\Psi^{RA})-I_c^{R\to B}(\sigma^{RB})\\
  &=S(\rho^A)-I_c(\rho^A,\mN)\\
  &\ge 0,
\end{split}
\end{equation}
where the non-negativity follows from the data-processing
inequality~\cite{data-proc}.

The following theorem (whose first part is in Ref.~\cite{schum-west}
and second part in Ref.~\cite{barnischu}) is exactly what we were
searching for

\begin{theorem} Let $\rho^A$ be the input state for a channel
  $\mN:A\to B$. Let $\delta_c(\rho^A,\mN)$ be the corresponding loss
  of coherent information. Then, there exists a recovering channel
  $\mR_{\rho,\mN}:B\to A$ such that
\begin{equation}\label{eq:direct}
  F_e(\rho^A,\mR_{\rho,\mN}\circ\mN)\ge 1-\sqrt{2\delta_c(\rho^A,\mN)}.
\end{equation}
Conversely, for every channel $\mR:B\to A$, it holds
\begin{equation}\label{eq:converse}
  \delta_c(\rho^A,\mN)\le\operatorname{g}(1-F_e(\rho^A,\mR\circ\mN)),
\end{equation}
where $\operatorname{g}(x)$ is an appropriate positive, continuous,
concave, monotonically increasing function such that $\lim_{x\to
  0}\operatorname{g}(x)=0$. In particular, for $x\le 1/2$, we can take
$\fg(x)=4x\log_2(d_A/x)$. $\blacksquare$
\end{theorem}
Notice that, in particular, we have
\begin{equation}\label{eq:miao}
\overline{F}_e(\rho^A,\mN)\ge1-\sqrt{2\delta_c(\rho^A,\mN)},
\end{equation}
 and
\begin{equation}\label{eq:miao2}
\delta_c(\rho^A,\mN) \le\fg(1-\overline{F}_e(\rho^A,\mN)),
\end{equation}
where $\overline{F}_e(\rho^A,\mN)$ was given in
Eq.~(\ref{eq:corr_ent_fid}).

The above theorem can be summarised by stating that the loss of
coherent information of an input pure state $|\Psi^{RA}\>$ due to a
channel $\mN$ acting on $A$ is small (that means
$\delta_c(\rho^A,\mN)$ close to zero) if and only if the channel $\mN$
can be approximately corrected on the support of $\rho^A$ (that means
$\overline{F}_e(\rho^A,\mN)$ close to one). This has been a very
important generalisation of the previous theorem appeared
Ref.~\cite{data-proc} concerning \emph{exact} channel correction,
namely $\mR\circ\mN=\id$ on the support of the input state $\rho^A$,
which turns out to be possible, as a corollary, if and only if
$\delta_c(\rho^A,\mN)=0$. Such a generalisation lies at the core of
some recent coding theorems for quantum channel capacity---see for
example Ref.~\cite{decoupling}.

Coherent information loss is now an extremely handy quantity to deal
with, easy to compute and providing sufficiently tight bounds on the
invertibility of the noise. However, coherent information still lacks
of some requirements we asked for in our original program. Indeed, we
would like to relate the degree of invertibility of a general quantum
noise to some function quantifying the loss of entanglement. In
fact, it is known that coherent information is not a satisfactory
measure of entanglement, and it is not straightforward to generalise
Theorem~1 to other entanglement measures loss. To find a relation
between noise invertibility and various entanglement measures will be
the aim of the next section.

\section{Entanglement loss(es)}

In the following we will focus on a widely studied family of possible
entanglement measures\footnote{This is the reason for the plural in
  the title.}, namely those which stem from von Neumann
entropy and analogous quantities. Among these measures, that often
gain an operational interpretation as the optimum asymptotic rate at
which a particular entanglement transformation can be done, we find,
e.~g. the \emph{distillable entanglement} $E_d$, the \emph{distillable
  key} $K_d$, the \emph{squashed entanglement} $E_{sq}$, the
\emph{relative entropy of entanglement} $E_r$, the \emph{entanglement
  cost} $E_c$, and the \emph{entanglement of formation} $E_f$, just to
mention some of them (for an accurate review of definitions and
properties of a large class of entanglement measures see
Ref.~\cite{christhesis} and references therein). In particular, in the
following we will explicitly call for the entanglement of formation,
which is defined as~\cite{eof}
\begin{equation}
E_f(\tau^{AB}):=\min\sum_ip_iE(\phi_i^{AB}),
\end{equation}
where the minimum is taken over all possible ensemble decompositions
$\tau^{AB}=\sum_ip_i\phi^{AB}_i$, for pure $\phi_i$'s, and where
$E(\phi^{AB}):=S(\Tr_B[\phi^{AB}])$, is the so-called \emph{entropy of
  pure-state entanglement}. Here we refrain from provide even a short
review of the other entropic-like entanglement measures we mentioned,
which would be far beyond the scope of the present contribution. The
interested reader is directed to Refs.~\cite{christhesis}
and~\cite{hayashi}. For our purposes, we are content with recalling
that, given a bipartite state $\tau^{AB}$, the following inequalities
hold
\begin{equation}\label{eq:bounds}
\begin{split}
  E_d(\tau^{AB})&\le K_d(\tau^{AB})\le E_{sq}(\tau^{AB})\le I^{A:B}(\tau^{AB})/2,\\
  K_d(\tau^{AB})&\le E_r(\tau^{AB})\le E_f(\tau^{AB}),\\
  E_{sq}(\tau^{AB})&\le E_c(\tau^{AB})\le E_f(\tau^{AB}),
\end{split}
\end{equation}
where $I^{A:B}(\tau^{AB}):=S(\tau^A)+S(\tau^B)-S(\tau^{AB})$ is the
\emph{quantum mutual information}. Moreover
\begin{equation}\label{eq:hashing}
\begin{split}
\max\{I_c^{A\to B}(\tau^{AB}),0\}&\le E_d(\tau^{AB}),\\
E_f(\tau^{AB})&\le\min\{S(\tau^A),S(\tau^B)\}.
\end{split}
\end{equation}
Notice that it is commonly found that
\begin{equation}
E_d(\tau^{AB})\ll E_{sq}(\tau^{AB})\ll E_f(\tau^{AB}),
\end{equation}
and, as dimensions of subsystems $A$ and $B$ increase, a mixed state
picked up at random in the convex set of mixed bipartite states almost
certainly (that is, with probability approaching one exponentially
fast in the dimension) displays an even more dramatic
separation~\cite{generic_ent}
\begin{equation}\label{eq:extreme-bounds}
E_d(\tau^{AB})\approx 0,\qquad E_f(\tau^{AB})\approx\min\{S(\tau^A),S(\tau^B)\}.
\end{equation}

Our motivation is to work out a result analogous to Theorem~1, where,
instead of the coherent information loss $\delta_c(\rho^A,\mN)$
introduced in Eq.~(\ref{eq:coherent-info-loss}), we would like to use
some other entanglement measure loss
\begin{equation}
\delta_x(\rho^A,\mN):=S(\rho^A)-E_x(\sigma^{RB}),
\end{equation}
where the letter ``$x$'' could stand, for example, for ``$sq$''
(squashed entanglement loss) or ``$f$'' (entanglement of formation
loss).

Already at a first glance, we can already say that, thanks to
Eqs.~(\ref{eq:bounds}-\ref{eq:hashing}), the second part of Theorem~1
can be extended to other entanglement loss measures, that is
\begin{equation}\label{eq:converse2}
  \delta_x(\rho^A,\mN)\le\delta_c(\rho^A,\mN)\le\fg(1-F_e(\rho^A,\mR\circ\mN)),
\end{equation}
for every channel $\mR:B\to A$. Instead, the generalisation of the
first part of Theorem~1 is not straightforward: because of the typical
entanglement behaviour summarised in Eq.~(\ref{eq:extreme-bounds}), we
could easily have, for example, a channel causing a \emph{vanishingly
  small} entanglement of formation loss with, at the same time, a
relatively \emph{severe} coherent information loss.

Still, the following argument suggests that \emph{there must be} an
analogous of Eq.~(\ref{eq:miao}) for alternative entanglement losses:
In fact, when evaluated on pure states, all mentioned entanglement
measures coincide with the entropy of pure-state
entanglement. Moreover, many of these entanglement measures are known
to be continuous in the neighbourhood of pure states. This is
equivalent to the fact that, in the neighbourhood of pure states, they
have to be reciprocally boundable. Therefore, if the action of the
noise $\mN$ is ``sufficiently gentle'' and the output state
$\sigma^{RB}$ exhibits an entanglement structure which is
``sufficiently close'' to pure-state entanglement\footnote{Notice that
  this is not equivalent to the state $\rho^{RB}$ itself being pure. A
  trivial example of a mixed state with pure-state entanglement
  structure is given by $\rho^{RB}=\Psi^{RB_1}\otimes\rho^{B_2}$,
  where $B_1$ and $B_2$ are two subsystems of $B$.}, then it should be
possible to write the analogous of Eq.~(\ref{eq:miao}) in terms of
$\delta_{sq}(\rho^A,\mN)$ or $\delta_f(\rho^A,\mN)$, for example, as
well. The problem is to explicitly write down such analogous formula.

In Ref.~\cite{buscemi}, the interested reader can find the proof of
the following theorem

\begin{theorem}
  Let $\rho^A$ be the input state for a channel $\mN:A\to B$. Let
  $\delta_{sq}(\rho^A,\mN)$ and $\delta_{f}(\rho^A,\mN)$ be the
  corresponding losses of squashed entanglement and entanglement of
  formation, respectively. Then
\begin{equation}\label{eq:direct1}
  \overline{F}_e(\rho^A,\mN)\ge 1-2\sqrt{\delta_{sq}(\rho^A,\mN)},
\end{equation}
and
\begin{equation}\label{eq:direct2}
  \overline{F}_e(\rho^A,\mN)\ge 1-\sqrt{2(2d_Ad_B-1)^2\delta_f(\rho^A,\mN)}.\ \blacksquare
\end{equation}
\end{theorem}
Notice the large numerical factor, depending on the dimensions of the
underlying subsystems, in front of the entanglement of formation loss:
this feature is reminiscent of the previously mentioned
irreversibility gap between distillable entanglement and entanglement
of formation, and makes it possible the situation where the noise
causes a vanishingly (in the dimensions) small entanglement of
formation loss, even though its action is extremely dissipative with
respect to the loss of coherent information. On the contrary, the loss
of squashed entanglement seems to be an efficient indicator of
irreversibility, almost as good as the coherent information loss---in
fact, only an extra constant factor of $\sqrt{2}$ appears in
Eq.~(\ref{eq:direct1}) with respect to Eq.~(\ref{eq:miao})---; on the
other hand, it is symmetric under the exchange of the input system
with the output system, a property that does not hold for the coherent
information loss. Summarising this section, the important thing is
that there always exist a threshold (which is strictly positive for
finite dimensional systems) below which all entanglement losses become
equivalent, in the sense that they can be reciprocally bounded (it is
noteworthy that, in the case of squashed entanglement loss and
coherent information loss, we can have dimension-independent bounds,
which is a desirable property when dealing with quantum channels
alone, see Section~5 below).

\subsection{Distillable entanglement vs entanglement of formation}

It is interesting now to forget for a moment about the channel $\mN$
itself, and see what Eqs.~(\ref{eq:miao}),~(\ref{eq:direct1}),
and~(\ref{eq:direct2}) mean in terms of a given bipartite mixed state
only. First of all, notice that, for every mixed state $\tau^{AB}$,
there exist two pure states, $|\phi^{AA'}\>$ and $|\psi^{B'B}\>$, and
two channels, $\mN:A'\to B$ and $\mathcal{M}:B'\to A$, such that
$(\id^A\otimes\mN^{A'})(\phi^{AA'})=\tau^{AB}$ and
$(\mathcal{M}^{B'}\otimes\id^B)(\psi^{B'B})=\tau^{AB}$.

Now, for a given state $\tau^{AB}$, let us define
\begin{equation}
  \delta_c^{A\to B}(\tau^{AB}):=S(\tau^A)-I_c^{A\to B}(\tau^{AB}),
\end{equation} 
and
\begin{equation}
  \delta_x^{A\to B}(\tau^{AB}):=S(\tau^A)-E_x(\tau^{AB}),
\end{equation}
where the letter $x$ is used as before\footnote{The analogous
  quantities $\delta^{B\to A}_x(\tau^{AB})$ are defined in the same
  way, by simply exchanging subsystems labels, as $\delta^{B\to
    A}_x(\tau^{AB}):=S(\tau^B)-E_x^{B\to A}(\tau^{AB})$.}. Then,
Theorems~1 and~2 tell us that there exist channels $\mR:B\to A'$ and
$\mathcal{T}:A\to B'$, and two pure states, $|\tilde{\phi}^{AA'}\>$
and $|\tilde{\psi}^{B'B}\>$, with $\Tr_{A'}[\tilde\phi^{AA'}]=\tau^A$
and $\Tr_{B'}[\tilde\psi^{B'B}]=\tau^B$, such that
\begin{equation}
\begin{split}
  \<\tilde\phi^{AA'}|(\id^A\otimes\mR^B)(\tau^{AB})|\tilde\phi^{AA'}\>&\ge 1-\sqrt{2K_x\delta_x^{A\to B}(\tau^{AB})},\\
  \<\tilde\psi^{B'B}|(\mathcal{T}^A\otimes\id^B)(\tau^{AB})|\tilde\psi^{B'B}\>&\ge
  1-\sqrt{2K_x\delta_x^{B\to A}(\tau^{AB})},
\end{split}
\end{equation}
where $K_c=1$, $K_{sq}=2$, and $K_f=(2d_Ad_B-1)^2$. In a sense, either
$\delta^{A\to B}_x(\tau^{AB})$ or $\delta^{B\to A}_x(\tau^{AB})$ being
small\footnote{That means $\delta_x(\tau^{AB})\ll (2K_x)^{-1}$.}, it
means that the entanglement present in the state $\tau^{AB}$ is
basically pure-state entanglement, even if $\tau^{AB}$ is itself a
mixed state. This is the reason for which we can establish a
quantitative relation between typically inequivalent entanglement
measures, as the following corollary of Theorems~1 and~2 clearly
states~\cite{buscemi}
\begin{corollary}
  For an arbitrary bipartite mixed state $\tau^{AB}$, with
  $S(\tau^A)\le S(\tau^B)$, the following inequality holds
\begin{equation}\label{eq:gap}
  \delta_c^{A\to B}(\tau^{AB})\le\fg\left(\sqrt{2(2d_Ad_B-1)^2\delta_f^{A\to B}(\tau^{AB})}\right),
\end{equation}
where $\fg(x)$ is a function as in Eq.~(\ref{eq:converse}) in
Theorem~1. $\blacksquare$
\end{corollary}
This corollary is in a sense the quantitative version of the intuitive
argument given before Theorem~2, and it represents a first attempt in
complementing the findings of Ref.~\cite{generic_ent}, summarised in
Eq.~(\ref{eq:extreme-bounds}).
\begin{figure}[h,c]
\begin{center}
  \includegraphics[width=10cm]{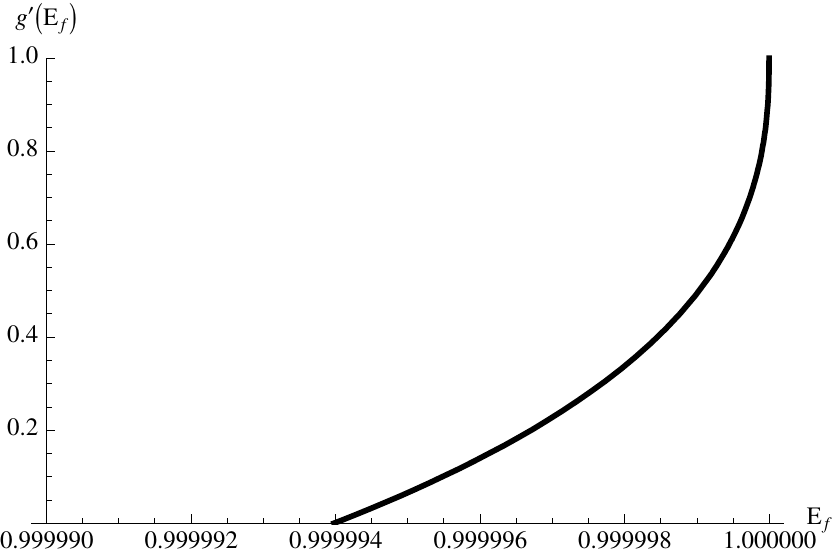}\end{center}
\caption{The plot (axes are normalised so that $\log_2(3)\mapsto 1$)
  shows the behaviour, for a bipartite system of two qutrits, of the
  lower bound in Eq.~(\ref{eq:fgprime}) for coherent information as a
  function of entanglement of formation. Coherent information, and
  hence distillable entanglement, are bounded from below by the thick
  curve. Notice that $E_f$ has to be extraordinarily close to its
  maximum value in order to have a non trivial bound from
  Eq.~(\ref{eq:fgprime}). This fact suggests that the bound itself
  could be improved.}
  \label{fig:2}
\end{figure}
It is also possible to invert Eq.~(\ref{eq:gap}) and obtain a
function $\fg'(E_f)$ such that
\begin{equation}\label{eq:fgprime}
  \fg'(E_f(\tau^{AB}))\le I_c^{A\to B}(\tau^{AB})\le E_d(\tau^{AB}),
\end{equation}
for all bipartite state $\tau^{AB}\in\states(A\otimes B)$. The plot of
$\fg'(E_f)$ is given in Fig.~\ref{fig:2} for $d_A=d_B=3$ (for qubits
every entangled state is also distillable), for a state for which
$\tau^A=\tau^B=\openone/3$. The plotted curve displays the typical
behaviour of the bound~(\ref{eq:fgprime}). Notice from
Fig.~\ref{fig:2} that entanglement of formation has to be extremely
close to its maximum attainable value in order to obtain a non trivial
bound from Eq.~(\ref{eq:fgprime}). This is a strong evidence that the
bound itself could probably be improved. Nonetheless, we believe that
such an improvement, if possible, would only make smaller some
(unimportant) constants which are independent of the dimension, while
leaving the leading order of dependence on $d=d_Ad_B$ in the right
hand side of Eq.~(\ref{eq:gap}) untouched.

\section{Overall channel invertibility: relations between entanglement
  losses and other measures of invertibility}

The previous analysis, following Ref.~\cite{buscemi}, was done in
order to quantify the invertibility of a noisy evolution with respect
to \emph{a given} input state $\rho^A$. In this section, we want to
derive quantities characterising the ``overall'' invertibility of a
given channel. In other words, we would like to get rid of the
explicit dependence on the input state and obtain the analogous of
Eqs.~(\ref{eq:miao}), (\ref{eq:miao2}), (\ref{eq:direct1}),
and~(\ref{eq:direct2}) as functions of the channel $\mN$ only.

Intuitively, to do this, we should quantify how close the corrected
channel $\mR\circ\mN$ can be to the noiseless channel $\id$, for all
possible corrections $\mR$. However, in doing this, we have to be very
careful about which channel distance function we adopt in order to
measure ``closeness''. A safe choice consists in using the distance
induced by the so-called \emph{norm of complete boundedness}, for
short \emph{cb-norm}, defined as
\begin{equation}\label{eq:def-cb}
  \N{\mN}_{cb}:=\sup_n\N{\id_n\otimes\mN}_{\infty},
\end{equation}
where $\id_n$ is the identity channel on $n\times n$ density matrices,
and
\begin{equation}
  \N{\mN}_{\infty}:=\sup_{\rho\ge0:\Tr[\rho]\le 1}\Tr\left[\,|\mN(\rho)|\,\right].
\end{equation}
(We put the absolute value inside the trace because in literature one
often deals also with non completely positive maps, so that the
extension $\id_n\otimes\mN$ can be non positive.) Notice, that, in
general, $\N{\mN}_{cb}\ge\N{\mN}_{\infty}$, and the two norms can be
inequivalent~\cite{dennis2}. A part of the rather technical definition
of cb-norm (the extension in Eq.~(\ref{eq:def-cb}) is necessary,
basically for the same reasons for which we usually consider complete
positivity instead of the simple positivity), we will be content with
knowing that, for channels, $\N{\mN}_{cb}=1$ and
$\N{\mN_1\otimes\mN_2}_{cb} =\N{\mN_1}_{cb}\N{\mN_2}_{cb}$, and that
the following theorem holds~\cite{dennis}
\begin{theorem}
  Let $\mN:A\to A$ be a channel, with $d_A<\infty$. Then
\begin{equation}
\begin{split}
  1-\inf_{\rho^A}F_e(\rho^A,\mN)&\le4\sqrt{\N{\mN-\id}_{cb}}\\
  \N{\mN-\id}_{cb}&\le4\sqrt{1-\inf_{\rho^A}F_e(\rho^A,\mN)},
\end{split}
\end{equation}
where the infimum of the entanglement fidelity is done over all
normalised states $\rho^A\in\states(A)$. $\blacksquare$
\end{theorem}
It is then natural to define a cb-norm--based measure of the overall
invertibility of a given channel $\mN:A\to B$ as
\begin{equation}\label{eq:cb-invert}
  \qcb(\mN):=\inf_{\mR}\N{\mR\circ\mN-\id}_{cb},
\end{equation}
with the infimum taken over all possible correcting channels $\mR:B\to
A$.

For a moment, let us now go back to the other functions we introduced
before. We will be able to relate them, in some cases with dimension
independent bounds, to the cb-norm--based invertibility
$\qcb(\mN)$. Given the loss function $\delta_x(\rho^A,\mN)$, where
$x\in\{c,sq,f\}$ is used to denote the coherent information loss, the
squashed entanglement loss, and the entanglement of formation loss,
respectively, we define the following quantity
\begin{equation}
\Delta_x(\mN):=\sup_{\rho^A}\delta_x(\rho^A,\mN),
\end{equation}
where the supremum is taken over all possible input states
$\rho^A$. Analogously, from Eq.~(\ref{eq:corr_ent_fid}), let us define
\begin{equation}
\begin{split}
\Phi(\mN)&:=\inf_{\rho^A}\overline{F}_e(\rho^A,\mN)\\
&=\inf_{\rho^A}\sup_{\mR_{\rho,\mN}}F_e(\rho^A,\mR_{\rho,\mN}\circ\mN).
\end{split}
\end{equation}
Such quantities are now functions of the channel only, and we want to
understand how well $\Delta_x(\mN)$ and $\Phi(\mN)$ capture the
``overall'' invertibility of a channel.

First of all, let us understand how they are related. Let
$\overline{\rho}$ be the state for which $\Delta_x(\mN)$ is
achieved. Then,
\begin{equation}
\begin{split}
\Delta_x(\mN)&=\delta_x(\overline{\rho},\mN)\\
&\le\fg(1-\overline{F}_e(\overline{\rho},\mN))\\
&\le\fg(1-\Phi(\mN)).
\end{split}
\end{equation}
On the other hand, let $\Phi(\mN)$ be achieved with
$\underline{\rho}$. Then,
\begin{equation}
\begin{split}
\Phi(\mN)&=\overline{F}_e(\underline{\rho},\mN)\\
&\ge 1-\sqrt{2K_x\delta_x(\underline{\rho},\mN)}\\
&\ge 1-\sqrt{2K_x\Delta_x(\mN)},
\end{split}
\end{equation}
where, as usual, $K_c=1$, $K_{sq}=2$, and $K_f=(2d_Ad_B-1)^2$.

We are now in position, thanks to Theorem~3, to show how $\qcb(\mN)$,
$\Delta_x(\mN)$, and $\Phi(\mN)$ are related to each other. Let the
value $\Phi(\mN)$ be achieved by the couple
$(\underline{\rho},\overline{\mR})$. Then,
\begin{equation}
\begin{split}
  \qcb(\mN)&\le\N{\overline{\mR}\circ\mN-\id}_{cb}\\
  &\le4\sqrt{1-\inf_{\rho^A}F_e(\rho^A,\overline{\mR}\circ\mN)}\\
  &=4\sqrt{1-F_e(\underline{\rho},\overline{\mR}\circ\mN)}\\
  &=4\sqrt{1-\Phi(\mN)}\\
  &\le4\sqrt[4]{2K_x\Delta_x(\mN)},
\end{split}
\end{equation}
where in the second line we used Theorem~3, since the channel
$\overline{\mR}\circ\mN$ has equal input and output spaces.

Conversely, let $\Delta_x(\mN)$ be achieved by
$\overline{\rho}$. Then, thanks to Eq.~(\ref{eq:converse2})
\begin{equation}
\begin{split}
\Delta_x(\mN)&\le\fg(1-F_e(\overline{\rho},\mR\circ\mN))\\
&\le\fg(1-\inf_{\rho^A}F_e(\rho^A,\mR\circ\mN)),
\end{split}
\end{equation}
for all channels $\mR:B\to A$. Let $\underline{\mR}$ be the channel
achieving the infimum in Eq.~(\ref{eq:cb-invert}). Then,
\begin{equation}
\begin{split}
\Delta_x(\mN)&\le\fg\left(1-\inf_{\rho^A}F_e(\rho^A,\underline{\mR}\circ\mN)\right)\\
&\le\fg\left(4\sqrt{\N{\underline{\mR}\circ\mN-\id}_{cb}}\right)\\
&=\fg\left(4\sqrt{\qcb(\mN)}\right).
\end{split}
\end{equation}
Summarising, we found that
\begin{equation}\label{eq:final-bounds}
\begin{split}
\Delta_x(\mN)&\le\fg\left(4\sqrt{\qcb(\mN)}\right)\\
\qcb(\mN)&\le4\sqrt[4]{2K_x\Delta_x(\mN)}.
\end{split}
\end{equation}
In the function $\fg(x)$, the dependence on the dimension $d$ is
present (see Theorem~1), however only inside a logarithm: this is not
bad, in view of coding theorems. The dependence on $d$ can instead be
dramatic in $K_f$; on the contrary, both $K_c$ and $K_{sq}$ are
independent on the dimension.

\appendix

\section*{Acknowledgements}

This work is funded by Japan Science and Technology Agency, through
the ERATO-SORST Project on Quantum Computation and Information. The
Author would like to thank in particular M~Hayashi, and K~Matsumoto
for interesting discussions and illuminating suggestions.

\end{document}